\documentclass[twocolumn]{aastex631}

\defcitealias{Lee2018}{L18} 

\usepackage{multirow} %% mlutirow in table 
\usepackage{kotex}
\usepackage{amsmath} 

\received{}
\revised{}
\accepted{October 16, 2024}

\submitjournal{ApJ}

\begin{document}

\title{Comparison of chemical compositions between the bright and faint red clumps for the metal-poor and metal-rich populations in the Milky Way bulge}

\correspondingauthor{Young-Wook Lee}
\email{ywlee2@yonsei.ac.kr}

\author[0000-0003-4364-6744]{Seungsoo Hong}
\affiliation{Department of Physics and Astronomy, Seoul National University, 1 Gwanak-ro, Gwanak-gu, Seoul 08826, Republic of Korea}
\affiliation{Center for Galaxy Evolution Research \& Department of Astronomy, Yonsei University, Seoul 03722, Republic of Korea}

\author[0000-0001-7277-7175]{Dongwook Lim}
\affiliation{Center for Galaxy Evolution Research \& Department of Astronomy, Yonsei University, Seoul 03722, Republic of Korea}

\author[0000-0002-2210-1238]{Young-Wook Lee}
\affiliation{Center for Galaxy Evolution Research \& Department of Astronomy, Yonsei University, Seoul 03722, Republic of Korea}



\begin{abstract}
We examined the double red clump (RC) observed in the Galactic bulge, interpreted as a difference in distance (“X-shaped bulge scenario”) or in chemical composition (“multiple population scenario”). To verify chemical differences between the RC groups, we performed low-resolution spectroscopy for RC and red giant branch (RGB) stars using Gemini-South/GMOS in three fields of the bulge, and collected diverse data from literature. We divided our sample stars not only into bright and faint RC groups, but also into bluer ([Fe/H] $<$ -0.1) and redder ([Fe/H] $>$ -0.1) groups following the recent $u$-band photometric studies. For the metal-poor stars, no statistically significant difference in CN index was detected between the bright and faint RC groups for all observed fields. However, we found, from cross-matching with high-resolution spectroscopic data, a sign of Na enhancement in the ‘metal-poor \& bright’ RC group compared to the ‘metal-poor \& faint’ group at $(l,b)=(-1^\circ,-8.5^\circ)$. When the contributions of the RGB stars on the RC regimes are taken into account, the Na abundance difference between genuine RCs would correspond to $\Delta$[Na/Fe]$\simeq$0.23 dex, similar to globular cluster (GCs) with multiple populations. In contrast, the metal-rich stars do not show chemical differences between the bright and faint RCs. It implies that the double RC observed in the metal-poor component of the bulge might be linked to the multiple populations originated from GC-like subsystem, whereas that of the metal-rich component would have produced by the X-shaped structure. Our results support the previous studies suggesting composite nature of the Milky Way bulge.
\end{abstract}

\keywords{Stellar populations(1622) --- Spectroscopy(1558) --- Red giant clump(1370) --- Galactic bulge(2041) --- Globular star clusters(656) --- Milky Way formation(1053)}


\section{Introduction} \label{sec:intro}

%%% Figure 1 %%%%%%%%%%%%%%%%%%%%%%%%%%%%%%%%%%%%%%%%%%%%%%%%%%%%%%%%
\begin{figure*}
\centering
   \includegraphics[width=0.8\textwidth]{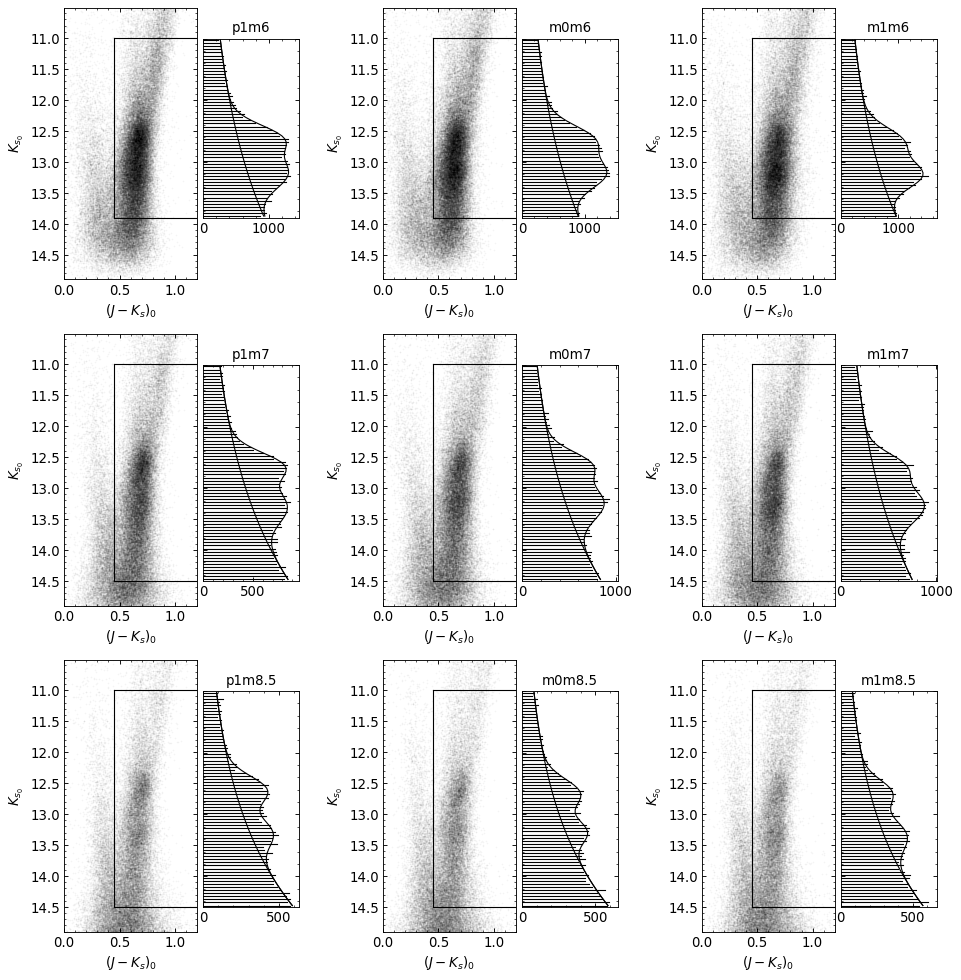} 
     \caption{$(J-K_s,K_s)_0$ CMDs and magnitude distributions of red giant branch (RGB) and RC stars for 9 fields toward the bulge at different combinations of Galactic longitude and latitude ($l=-1^\circ,0^\circ,+1^\circ$; $b=-8.5^\circ,-7^\circ,-6^\circ$). Photometric data are from 2MASS All-Sky Point Source Catalog (\citealt{Skrutskie2006}). The thick black lines in the CMD represent the color-magnitude selection boxes for the possible RGB+RC stars in the bulge to reduce contamination of the foreground stars. Exponential fits for RGB and gaussian fits for the double RC are displayed by black solid lines in the magnitude histograms. The meaning of field names is the $l$ and $b$ of each field, where the `p' and `m' are for the plus and minus values of $l$ and $b$: e.g., m1m8.5 denotes the field at $(l,b)=(-1^\circ,-8.5^\circ)$. Note that two peaks of the red clump are clearly observed, and the double peak feature changes with $l$ and $b$.}
     \label{Fig1}
\end{figure*}
%%%%%%%%%%%%%%%%%%%%%%%%%%%%%%%%%%%%%%%%%%%%%%%%%%%%%%%%%%%%%%%%%%%%%

In 2010, the presence of two red clumps (RCs) was discovered in the color-magnitude diagram (CMD) of stars in the high-latitude fields of the Milky Way (MW) bulge from the wide-field photometric surveys in near-infrared (NIR) and optical wavelength regions (\citealt{McWilliam2010,Nataf2010}). The magnitude difference between the bright RC (bRC) and the faint RC (fRC) is $\sim$0.5 mag in $K_s$ band in the high-latitude ($\vert b \vert$ $\sim$~$8^\circ$) fields of the bulge. The almost identical $J-K_s$ and $V-I$ colors of the bRC and fRC regardless of galactic longitude and latitude suggested that the metallicity difference between the two RC groups is indiscernible (\citealt{McWilliam2010,Nataf2010}), which is strengthened by the follow-up spectroscopic observations (\citealt{DePropris2011,Uttenthaler2012}). Since the RC stars have widely been used as a distance indicator (\citealt{Cannon1970,Stanek1994,Girardi2016}), the X-shaped bulge model has been suggested to explain the double RC phenomenon (\citealt{McWilliam2010,Nataf2010,Li2012,Wegg2013}). In this model, the X-shaped structure could be formed from the buckling instability of a bar (\citealt{Combes1990,Raha1991,Debattista2017}). Some difference in proper motion between the two RC groups reported by \citeauthor{Sanders2019} (\citeyear{Sanders2019}; see also \citealt{Clarke2019}) supports the X-shaped structure of the bulge, although there is a room for alternative explanations for the observed difference (see, e.g., \citealt{Lee2019}). Therefore, this model suggests that the MW bulge is a pseudo bulge formed from secular evolution. It is also worth to note that peanut or X-shaped bulges are occasionally observed in external galaxies (\citealt{Bureau2006,Gonzalez2016}).

A completely different interpretation for the origin of double RC feature, however, has been suggested by \citeauthor{Lee2015} (\citeyear{Lee2015}; see also, \citealt{Lee2016,Joo2017}) based on the multiple stellar population phenomenon, which is generally observed in globular clusters (GCs; see, e.g., \citealt{Bastian2018}). This model is based on the standard stellar evolution model in which He-enhanced second- and later-generation (G2+) stars are placed at the bRC in the metal-rich regime like the bulge (see Figure~1 of \citealt{Lee2019}). Therefore, the bRC in this model consists of G2+ stars enhanced in He, N, and Na abundances, while the fRC is representing first-generation (G1) stars with normal chemical abundances. As GC is known to be a unique environment that produces He-enhanced G2+ and He-normal G1 populations simultaneously (\citealt{Gratton2012,Renzini2015,Bastian2018}), the presence of two RCs in this model implies that GC-like systems have contributed to the formation of the bulge as building blocks. Recent spectroscopic studies reported a statistically significant difference in chemical abundances between the bRC and fRC as an evidence of the multiple population scenario \citep{Lee2018,Lim2021a}. Especially, based on the [Fe/H] difference between the two RC groups and similar chemical patterns as observed in the metal-rich bulge GC Terzan 5 showing double RC feature, \citet{Lim2021a} suspected Terzan 5-like stellar systems as main building blocks of the formation of the MW bulge (see also, \citealt{Ferraro2009,Ferraro2016}). A bulge GC Liller 1, whose structural parameters are very similar to those of Terzan 5 (\citealt{Saracino2015}), has recently been revealed to host multiple stellar populations, supporting the scenario that assembly of Terzan 5-like stellar systems have contributed to the formation of the MW bulge (\citealt{Ferraro2021}).

One of the important aspects of the double RC phenomenon is the variation of the double peak feature depending on Galactic longitude ($l$) and latitude ($b$) (See Figure~\ref{Fig1}). The magnitude difference between the bRC and fRC is $\sim$0.5 mag in high $b$ ($\vert b \vert$ $\sim$~$8^\circ$) fields, but the separation becomes less clear toward the lower $b$ fields. In addition, the relative strength of the bRC to fRC increases towards positive $l$, while it decreases in negative $l$. In the X-shaped bulge scenario, the tilt of the X-shaped structure with respect to the line-of-sight same as the bar component can explain the $l$ and $b$ dependence of the double RC feature (see, e.g., \citealt{McWilliam2010,Wegg2013}). The multiple population model, however, can also naturally explain the $l$ and $b$ dependence by the effect of tilted bar component embedded in the spheroidal bulge. The bar component, which places between the two RCs and fills the gap on the magnitude distribution, becomes more prominent at lower $b$ fields. Moreover, since the bar is placed at the near (far) side of the bulge toward positive (negative) $l$, the bar component gradually moves from bright to faint as the $l$ changes from positive to negative (see \citealt{Lee2015} and \citealt{Joo2017} for detailed model). Because of the different contributions from the bar and the bulge components in different fields, the multiple population model predicts a $l$ and $b$ dependence of the difference in chemical abundances between the bRC and fRC. However, this variation is yet to be confirmed from spectroscopic observations.

In order to investigate the chemical abundance difference between the bright and faint RC groups as well as its variation with $l$ and $b$, we performed low-resolution spectroscopic observations for the RC and red gainat branch (RGB) stars in the three fields of the bulge at $(l,~b) = (-1^\circ,~-8.5^\circ), (-1^\circ,~-7^\circ), (+1^\circ,~-7^\circ)$. In addition, for more diverse analysis, we combined our data with previous high-resolution spectroscopic and photometric data. This paper is organized as follows. In Section~\ref{sec:observation}, we describe survey design, observation, and data reduction process. In Section~\ref{sec:PreviousData}, we introduce RC subgrouping using data from previous studies. Our main results are presented in Section~\ref{sec:result1} and Section~\ref{sec:result2&3}. Finally, discussion and conclusion for the origin of double RC and its implication for the formation of the MW bulge are given in Section~\ref{sec:Discussion&Conclusion}.

\section{Spectroscopic survey for the red clump stars in the bulge} \label{sec:observation}
\subsection{Survey design and target selection}
Our pilot observations using WFCCD at Las Campanas Observatory (LCO) 2.5m telescope (\citealt{Lee2018}, hereafter \citetalias{Lee2018}) reported the difference in CN band strength between the bRC and fRC at the `negative $l$, high $b$' (m1m8.5, see Figure~\ref{Fig1} for the field name codes) field. Since the [Fe/H] difference between the two groups had not been reported prior to \citet{Lim2021a}, the difference in CN band strength was attributed to the difference in N abundance as observed in GC (see, e.g., \citealt{Hong2021}). Therefore, \citetalias{Lee2018} suggested a possible enhancement in N abundance of bRC stars compared to the fRC stars, which is expected in the multiple population scenario.

In this study, we designed extended spectroscopic observations for the 3 different fields of the bulge (m1m7, p1m7, and m1m8.5) in order to confirm the chemical abundance differences between the two RCs at various fields, and to examine the dependence of the chemical abundance differences on $l$ and $b$. We performed low-resolution spectroscopy using Gemini Multi-Object Spectrograph on the Gemini-South 8.1 m telescope (hereafter GMOS-S). For checking consistency between our pilot and this observations, we revisited m1m8.5 field, which is observed in \citetalias{Lee2018}.

%%% Figure 2 %%%%%%%%%%%%%%%%%%%%%%%%%%%%%%%%%%%%%%%%%%%%%%%%%%%%%%%%
\begin{figure}
\centering
   \includegraphics[width=0.45\textwidth]{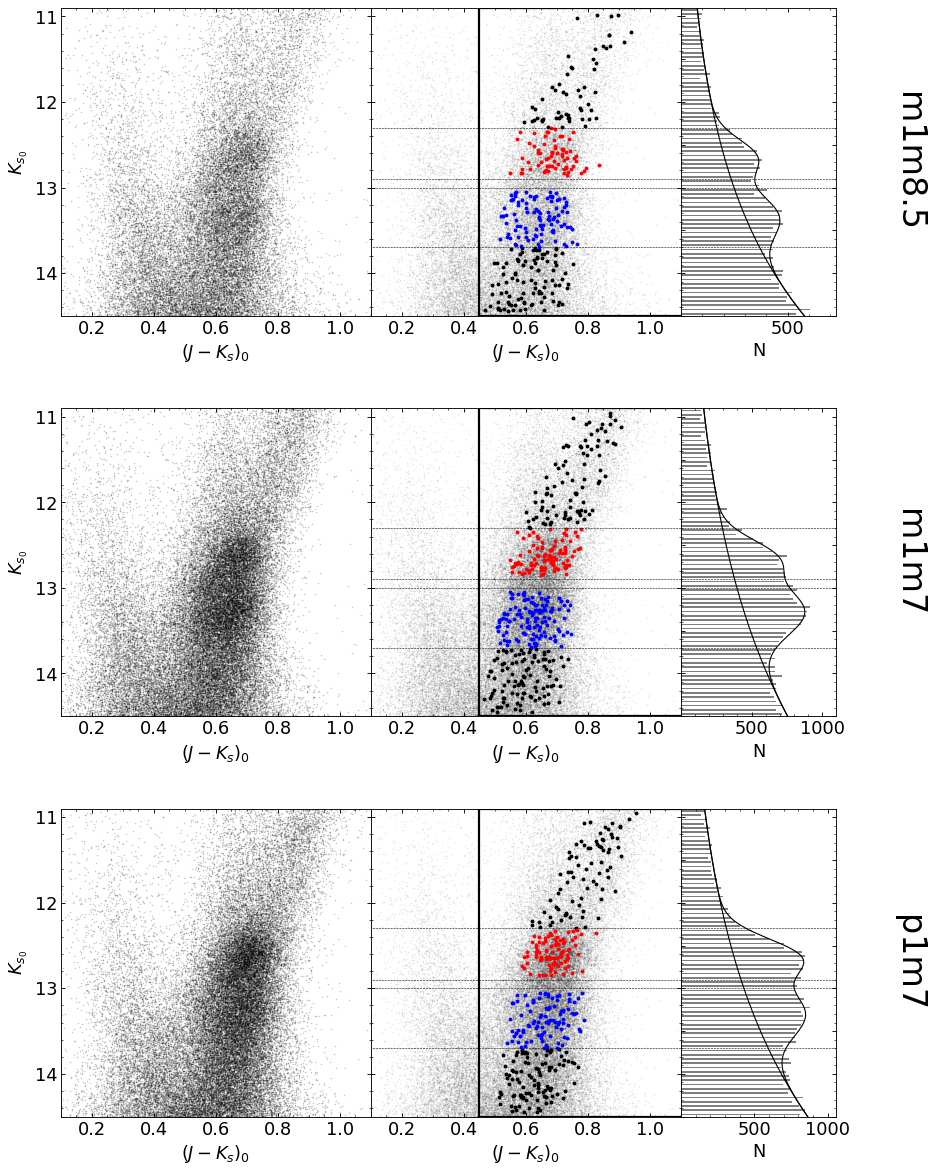} 
     \caption{$(J-K_s,K_s)_0$ CMDs for bulge stars within 1$^\circ$ radius from m1m8.5 ($l=-1^\circ$, $b=-8.5^\circ$; top), m1m7 ($l=-1^\circ$, $b=-7.0^\circ$; middle), and p1m7 ($l=+1^\circ$, $b=-7.0^\circ$; bottom) fields. The selected target stars for the observations with GMOS-S are indicated as red (bRC), blue (fRC), and black (RGB) circles. The thick black lines in the middle panels indicate the selection box for the bulge stars (see Figure~\ref{Fig1}). The magnitude histograms for RGB and RC stars with the fitted lines are also shown in the right panels.}
     \label{Fig2}
\end{figure}
%%%%%%%%%%%%%%%%%%%%%%%%%%%%%%%%%%%%%%%%%%%%%%%%%%%%%%%%%%%%%%%%%%%%%

%%% Table 1 %%%%%%%%%%%%%%%%%%%%%%%%%%%%%%%%%%%%%%%%%%%%%%%%%%%%%%%%
\begin{table}
%\scriptsize
\setlength{\tabcolsep}{8pt}
\caption{The numbers of observed stars}
\label{Tab1}
\centering
	\begin{tabular}{lllc}
		\hline\hline
		Field  & Obs. semester & Class & No. of stars \\ \hline
		m1m8.5 & 2020B/2021A   & bRGB  & 57           \\
		&               & bRC   & 70           \\
		&               & fRC   & 97           \\
		&               & fRGB  & 97           \\
		&               & total & 321          \\ \hline
		m1m7   & 2018B/2019A   & bRGB  & 89           \\
		&               & bRC   & 96           \\
		&               & fRC   & 126          \\
		&               & fRGB  & 102          \\
		&               & total & 413          \\ \hline
		p1m7   & 2019A/2020B   & bRGB  & 79           \\
		&               & bRC   & 88           \\
		&               & fRC   & 98           \\
		&               & fRGB  & 101          \\
		&               & total & 366          \\ \hline
	\end{tabular}
\end{table}
%%%%%%%%%%%%%%%%%%%%%%%%%%%%%%%%%%%%%%%%%%%%%%%%%%%%%%%%%%%%%%%%%%%%%

Target stars at each field were selected from $(J-K_s,K_s)_0$ CMDs, where the photometric data are from the Two Micron All Sky Survey (2MASS; \citealt{Skrutskie2006}) catalog. We defined bright RGB (bRGB; $10.9 < K_s \leq 12.3$), bRC ($12.3< K_s \leq 12.9$), fRC ($13.0 < K_s \leq 13.7$), and faint RGB (fRGB; $13.7 < K_s \leq 14.5$) groups based on the RGB + RC luminosity functions as shown in the right panels of Figure~\ref{Fig2}. We also set a gray zone ($12.9 < K_s \leq 13.0$), where the bRC and fRC are much overlapped. Stars placed in the gray zone were excluded from our target selection. In addition, we intended a comparable number of stars to be selected in each group. However, the number of stars in bRGB group is somewhat smaller than that in other groups, which is due to the limited number of bright stars at a given field-of-view (FOV). The selected target stars at m1m8.5, m1m7, and p1m7 fields are shown in Figure~\ref{Fig2}. The numbers of observed stars in each field are summarized in Table~\ref{Tab1}.

Slit masks for multi-object spectroscopy were designed using $GMMPS$, Gemini's mask making software. Typically, each mask includes $\sim$30 slits of 1$\arcsec$.0 width, and also contains three or four slit boxes for acquisition stars used to align the mask on sky. We note that slits for both faint and bright stars were made in the same mask in order to minimize any systematic bias between the fRC and bRC targets. 

\subsection{Observations and data reduction}
The spectral data were collected with GMOS-S during four observing runs in the 2018B, 2019A, 2020B, and 2021A semesters. These observations were carried out as part of the K-GMT science program under Program IDs: GS-2018B-Q-211, GS-2019A-Q-118, GS-2019A-Q-223, GS-2020B-Q-115, and GS-2021A-Q-215. The GMOS-S detector consists of three $2048 \times 4176$ Hamamatsu chips in a row with narrow 91 pixel gaps between them, and provides a 5.5 square arcmin FOV. We used the B600$-$G5323 grating with a central wavelength (CWL) of $\sim$4300 {\AA}, which provides a resolving power of $R$~$\sim$~1700 with a wavelength coverage of 3700 $-$ 6300 \AA, safely including the main spectral bands of CN absorptions near 3900 {\AA} and 4100 \AA, as well as CH near 4300 \AA, and Ca II H\&K lines. For each mask, spectra were obtained at two or three different CWLs with 100 {\AA} shifts, in order to fill the interchip gaps. At a given CWL, we took at least two consecutive 450s exposures for the science frame, followed by Quartz Halogen flat and CuAr arc exposures with the same observational settings. The integrated exposure time of the science frame for each mask is $\sim$2700 seconds in average, providing signal-to-noise ratio (S/N) $\sim$ 20 at 4000 \AA.

The science frames were reduced using the Gemini GMOS IRAF Package in a standard manner, including bias subtraction, flat fielding, wavelength calibration, and quantum efficiency corrections. We combined sequential images at each CWL to remove cosmic rays. From the combined images, spectra of each star were traced, sky-subtracted, and then extracted. The spectra covering full wavelength range were obtained by combining spectra taken at different CWLs. The full spectra were shifted to the rest-frame by correcting radial velocities which were measured using $xcsao$ task in IRAF $rvsao$ package. Finally, we performed careful visual inspections for every single spectrum to exclude spectra with unreasonable features in the analysis.

%%% Figure 3 %%%%%%%%%%%%%%%%%%%%%%%%%%%%%%%%%%%%%%%%%%%%%%%%%%%%%%%%
\begin{figure}
\centering
   \includegraphics[width=0.45\textwidth]{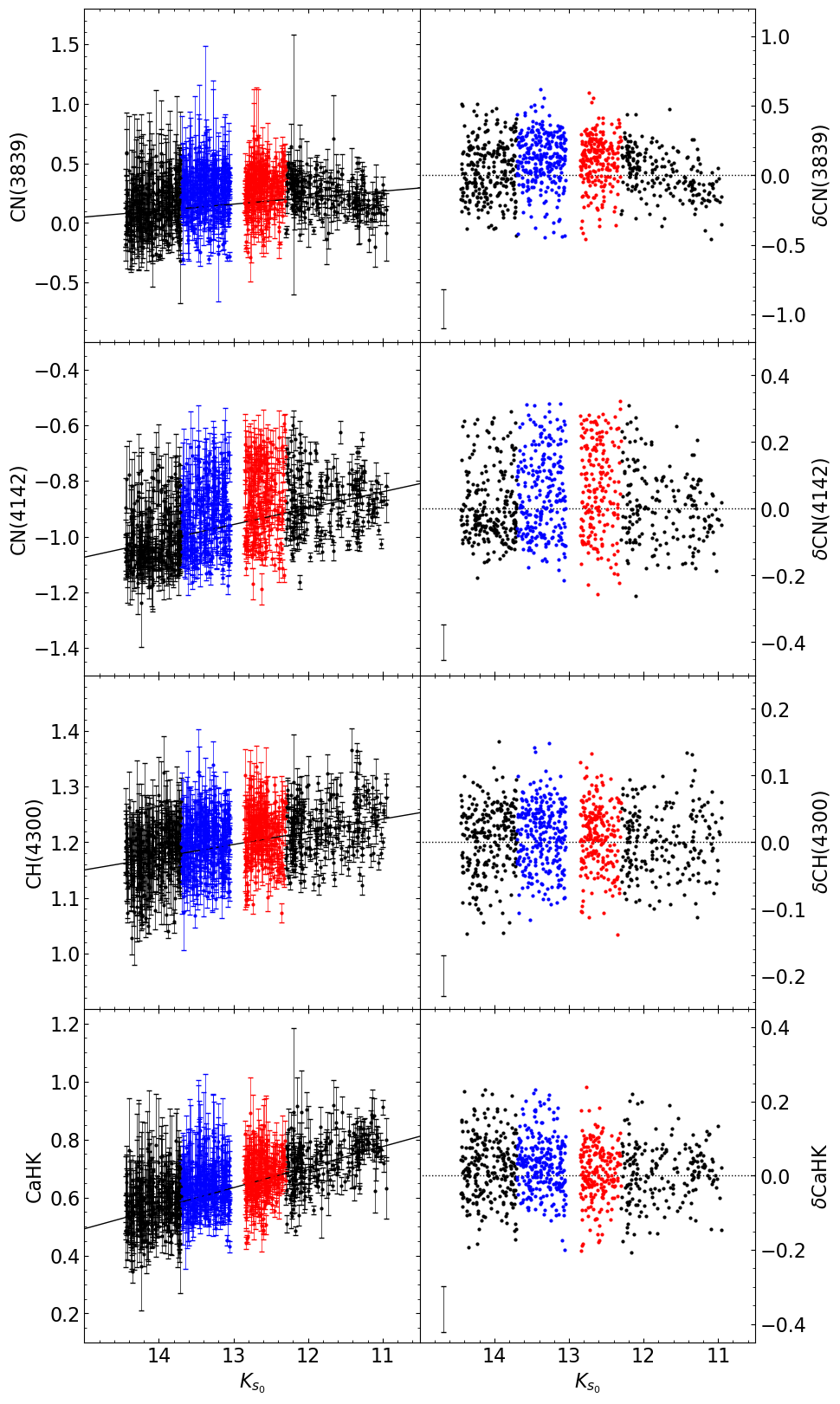} 
     \caption{
     Left panels: Measured spectral indices as functions of $K_s$ magnitude for RGB and RC stars in the bulge. The bRC, fRC, and RGB stars are indicated with the same color coding as Figure~\ref{Fig2}. Least-square fitting lines (black solid lines), which are used to derive $\delta$-indices, were obtained using stars only in the RGB regimes. 
     Right panels: $\delta$-indices plotted against $K_s$ magnitude. The vertical bars in the lower left corner indicate the typical measurement error for each index. Note that stars in the bRC and bright RGB regimes severely suffer from saturation in CN(3839) index.
     }
     \label{Fig3}
\end{figure}
%%%%%%%%%%%%%%%%%%%%%%%%%%%%%%%%%%%%%%%%%%%%%%%%%%%%%%%%%%%%%%%%%%%%%

%%% Figure 4 %%%%%%%%%%%%%%%%%%%%%%%%%%%%%%%%%%%%%%%%%%%%%%%%%%%%%%%%
\begin{figure}
\centering
   \includegraphics[width=0.45\textwidth]{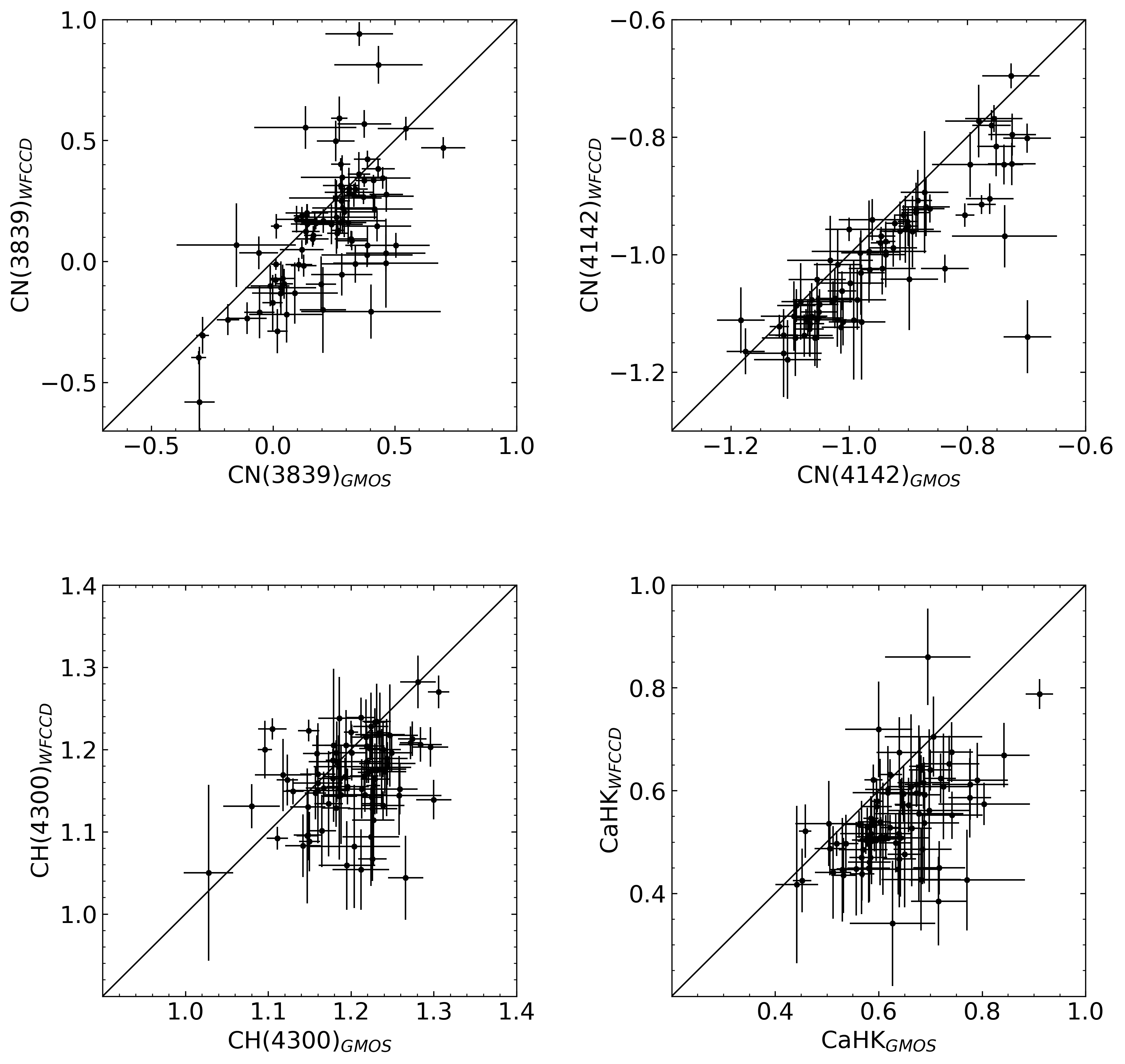} 
     \caption{Comparison of spectral indices measured from GMOS-S spectra with those measured in \citetalias{Lee2018} for the common sample stars. The solid line is for the one-to-one relation. Spectral indices measured from the two different instruments show good correlations.
     }
     \label{Fig4}
\end{figure}
%%%%%%%%%%%%%%%%%%%%%%%%%%%%%%%%%%%%%%%%%%%%%%%%%%%%%%%%%%%%%%%%%%%%%

\subsection{Line indices measurements} \label{indexmeasurement}
We measured CN(3839), CN(4142), CH(4300), and CaHK indices and their errors for each star following the same procedures employed in \citetalias{Lee2018}. The meaning of the indices is the ratio of absorption strength to the nearby continuum, and the Poisson statistics were used in the error estimation. The $\delta$-indices were also measured in order to minimize the effects of effective temperature and surface gravity in the line strengths (\citealt{Norris1983,Harbeck2003,Lim2015}). The $\delta$-indices were calculated as the difference between the original CN, CH, and CaHK indices and the linear-fit lines for stars in the RGB regimes, excluding stars with low-quality spectra (S/N at 4000 {\AA} $<$ 5), or with large measurement errors (outside 3$\sigma$ of the error distribution), or with abnormal index values (outside 3$\sigma$ of the index distribution). The linear fitting was performed with the \textit{linmix} package (\citealt{Kelly2007}) in order to take into account the intrinsic dispersion of the indices. The measured spectral indices as functions of $K_s$ magnitude are shown in Figure~\ref{Fig3}. The CN(4142), CH(4300), and CaHK indices are broadly increased as the magnitude decreased. These trends are consistent with the results of the previous studies for the bulge (\citetalias{Lee2018}) and observed in GCs (\citealt{Lim2017,Hong2021}). The CN(3839) index, however, shows a non-linear trend. The CN(3839) indices of stars in the fRGB, fRC, and bRC regimes are increased as the magnitude decreased like other indices, while those of stars in the bRGB regime are clearly decreasing\footnote{The posterior distribution of the slope for the bRGB stars indicates that the probability of having a negative slope (i.e., an increasing trend of CN(3839) as magnitude decreases) is extremely low ($P_{(slope < 0)} < 1\times10^{-5}$).}. This trend is probably due to the saturation of the CN(3839) index at low temperature of bRGB stars (\citealt{Gerber2020}). Although the CN(4142) index is also saturated at bRGB regime, this effect appears to be negligible for stars in fRGB, fRC, and bRC regimes and defining least-square line. It is also reported that the CN-band at near $\sim$3900 {\AA} is firstly saturated at higher temperature compared to that at near $\sim$4100 {\AA} \citep{Gerber2020}. Therefore, we used CN(4142) index as the main CN index for the following analysis.

\section{Data from previous studies} \label{sec:PreviousData}
\subsection{Data cross-matching}
Due to the severe extinction toward the bulge field, the MW bulge has been mainly studied using optical to NIR band observations, such as Optical Gravitational Lensing Experiment (OGLE; \citealt{Udalski2002}), 2MASS (\citealt{Skrutskie2006}), Vista Variables in the Via Lactea (VVV; \citealt{Minniti2010}), and Apache Point Observatory Galactic Evolution Experiment (APOGEE; \citealt{Majewski2017}) until 2020. The new photometric study on the MW bulge, covered near-ultraviolet (NUV) band, become available from 2020, thanks to the Blanco DECam Bulge Survey (BDBS; \citealt{Rich2020,Johnson2020}).

For advanced analysis on the origin of double RC, we collected previous spectroscopic and photometric data, in both NIR and NUV bands, for our target fields. For spectroscopic data, we cross-matched our GMOS-S data with WFCCD data from \citetalias{Lee2018} and Michigan/Magellan Fiber System (M2FS) high-resolution spectroscopic data from \citet{Lim2021a}, using the \textit{Match Tables} and \textit{CDS X-Match} services provided by $TOPCAT$ software (\citealt{Taylor2005}). These matching results in common sample of 85 stars between GMOS-S and WFCCD observations, and that of 40 stars between GMOS-S and M2FS observations. In the case of APOGEE survey, however, no star is matched with our observation due to the different targeting fields and magnitudes.

For checking the consistency of the index measurements, we compared the spectral indices for $\sim$80 stars in m1m8.5 field, which are commonly observed in GMOS-S and WFCCD, in Figure~\ref{Fig4}. These measurements are generally in good agreement, although indices measured from GMOS-S are somewhat larger compared to those observed from WFCCD. The offset is probably due to the difference in instruments and/or reduction procedures. Note, however, that the small offset does not affect our analysis because the comparisons of spectral indices between the different RC groups were made with values obtained from the same instrument.

The photometric data from VVV (\citealt{Minniti2010}) and BDBS (\citealt{Johnson2022}) are also combined with our spectroscopic data. We further used $Gaia$ DR3 calalogue (\citealt{GaiaCollaboration2016,GaiaCollaboration2023}) to exclude foreground sources in our sample, so the stars within the range of $\bar\omega$ (parallax) $<0.4$ and $\bar\omega/\sigma_{\bar\omega}$ (relative parallax error) $< 5.0$ are only considered as the possible bulge stars. We note that similar criteria were employed by \citet{Lim2021b} and \citet{Marchetti2022} to remove obvious foreground contamination sources from the bulge stars. In addition to the parallax cuts, following \citet{Marchetti2022}, we selected stars with $\sigma_{\mu_{\alpha^\ast}}$ (uncertainty on the proper motion in right ascention) $<1.5 \ mas\ yr^{-1}$, $\sigma_{\mu_{\delta}}$ (uncertainty on the proper motion in declination) $<1.5 \ mas\ yr^{-1}$, RUWE (renormalized unit weight error) $<1.4$, and stars placed at inside 3$\sigma$ of the proper motion distributions in the direction of $l$ and $b$ in order to exclude stars with unreliable astrometric measurements. Consequently, about 35\% of GMOS sample stars are excluded by this astrometric criteria.

%%% Figure 5 %%%%%%%%%%%%%%%%%%%%%%%%%%%%%%%%%%%%%%%%%%%%%%%%%%%%%%%%
\begin{figure}
\centering
   \includegraphics[width=0.45\textwidth]{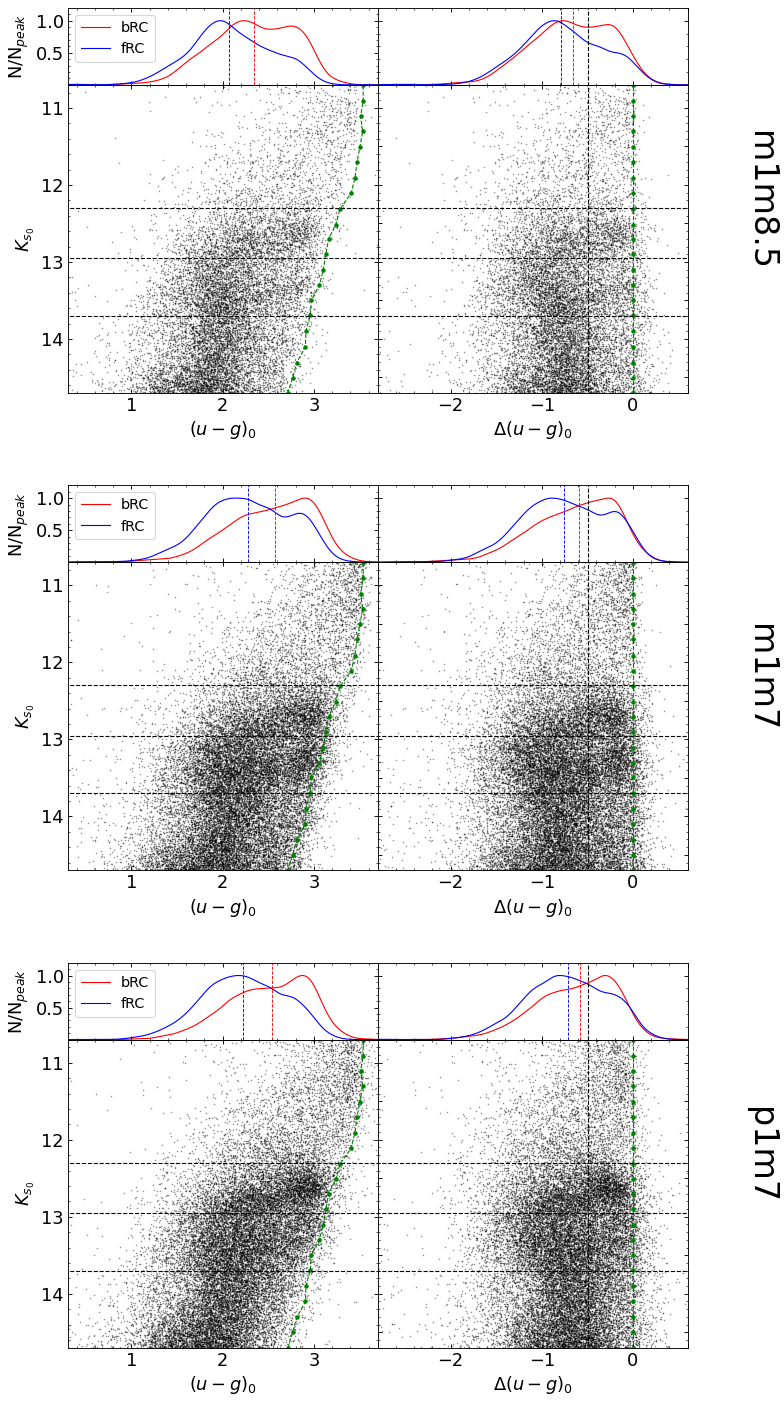} 
     \caption{Comparison of the $(u-g)_0$ and $\Delta(u-g)_0$ colors, and the color distributions of bRC and fRC. CMDs with $(u-g)_0$ and $\Delta(u-g)_0$ colors for stars in m1m8.5 (top), m1m7 (middle), and p1m7 (bottom) fields. The green line is the fiducial line adopted to verticalize the RGB (see text for details). $\Delta(u-g)_0$ is defined as the difference in $(u-g)_0$ of each star from the fiducial line. The horizontal and vertical black dashed lines represent the magnitude and color criteria dividing bright/faint and bluer/redder subgroups. The upper panels present comparisons of the $(u-g)_0$ and $\Delta(u-g)_0$ color distributions between the bRC and fRC, obtained with gaussian kernel density estimators (KDE). Vertical red and blue dashed lines indicate the median values of corresponded distributions.}
     \label{Fig5}
\end{figure}
%%%%%%%%%%%%%%%%%%%%%%%%%%%%%%%%%%%%%%%%%%%%%%%%%%%%%%%%%%%%%%%%%%%%%

%%% Figure 6 %%%%%%%%%%%%%%%%%%%%%%%%%%%%%%%%%%%%%%%%%%%%%%%%%%%%%%%%
\begin{figure}
\centering
   \includegraphics[width=0.45\textwidth]{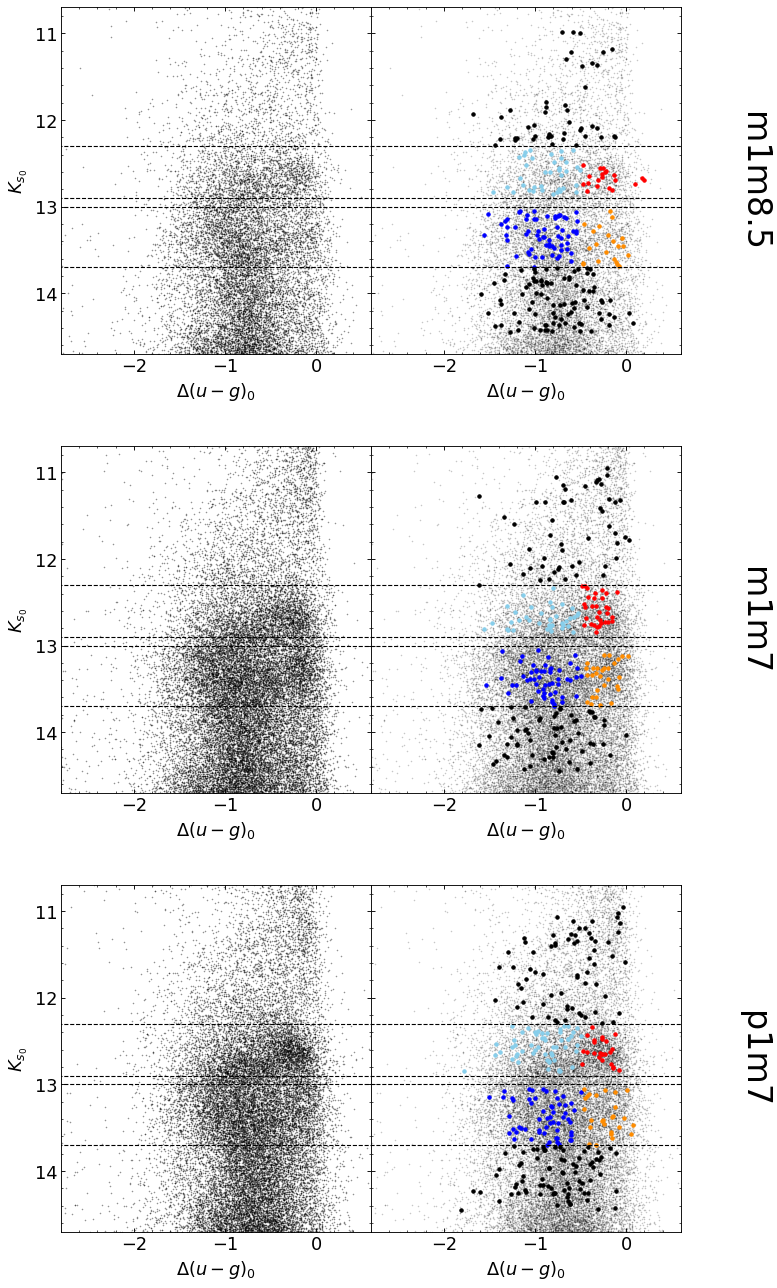} 
     \caption{
      The verticalized CMD with $\Delta(u-g)_0$ color for stars in the m1m8.5 (top), m1m7 (middle), and p1m7 (bottom) fields. Our sample stars observed with GMOS-S are indicated by colored circles, where blue, skyblue, orange, and red color represent bluer-fRC, bluer-bRC, redder-fRC, and redder-bRC, respectively (see Table~\ref{Tab2}).}
     \label{Fig6}
\end{figure}
%%%%%%%%%%%%%%%%%%%%%%%%%%%%%%%%%%%%%%%%%%%%%%%%%%%%%%%%%%%%%%%%%%%%%

\subsection{Red clump subgrouping based on BDBS $u$-band color} \label{subsec:RCsubgrouping}
Based on the BDBS photometric data, \citet{Johnson2020} and \citet{Lim2021b} found the bimodal distributions in $u$-band colors ($u-g$ and $u-i$) for both the bRC and fRC, which are not observed in NIR colors, such as $J-K_s$. It implies that each RC group can no longer be considered a single population as previously thought, but needs to be further divided into bluer and redder RCs (\citealt{Johnson2020,Lim2021b}). For instance, \citet{Lim2021b} divided stars in the RC regime into the bluer and redder groups using $(u-g)_0$ color, with $(u-g)_0<2.5$ for the bluer RC and $(u-g)_0\geq2.5$ for the redder RC.

However, applying the same $(u-g)_0$ criterion for both bRC and fRC to subdivide into bluer and redder groups would cause a selection bias because the RGB stars, which are totally overlapped with RC stars on the CMD of the bulge, show a trend of increasing color with decreasing magnitude. In order to minimize the bias caused by the RGB slope in RC subgrouping, we employed $\Delta(u-g)_0$ color index, which is defined as the difference in the $(u-g)_0$ of each star from the fiducial lines as shown in Figure~\ref{Fig5}. The fiducial lines were obtained by connecting points determined from the 96th percentile of the $(u-g)_0$ color distribution and the median $K_{s_0}$ magnitude in every 0.2 magnitude bins, following the procedure used in the study of multiple populations in GCs (\citealt{Milone2015}). As shown in the upper panels for each field in Figure~\ref{Fig5}, the bRC is systematically biased to be red in the $(u-g)_0$ color, but it can be minimized in the $\Delta(u-g)_0$ color index. Therefore, based on the $\Delta(u-g)_0$, we defined the bluer (stars with $\Delta(u-g)_0<-0.5$) and the redder (stars with $\Delta(u-g)_0\geq-0.5$) groups, which result in bluer-bRC, bluer-fRC, redder-bRC, and redder-fRC subgroups. The newly defined subgroups are summarized in Table~\ref{Tab2}, and the final spectroscopic sample stars in each group are indicated on the $(\Delta(u-g),K_s)_0$ CMDs in Figure~\ref{Fig6}.

The newly defined subgroups are summarized in Table~\ref{Tab2}. The final spectroscopic sample stars in each group are indicated on the $(\Delta(u-g),K_s)_0$ CMDs in Figure~\ref{Fig6}, and the measured indices and errors for the stars are listed in Table~\ref{Tab3}.

%%% Table 2 %%%%%%%%%%%%%%%%%%%%%%%%%%%%%%%%%%%%%%%%%%%%%%%%%%%%%%%%
\begin{table}
%\scriptsize
\setlength{\tabcolsep}{7pt}
\caption{Definition of red clump subgroups}
\centering                 
\begin{tabular}{ccc}
\hline\hline 
{Criteria}                      & \multicolumn{2}{c}{\textit{(color)}}                       \\ 
\cline{2-3}
{\textit{(magnitude)}}            & $\Delta(u-g)_0 < -0.5$    & $\Delta(u-g)_0 \geq -0.5$     \\
\hline
$K_{s_0}<12.3$                  & \multicolumn{2}{c}{bright RGB (bRGB)}                           \\
\hline
{$12.3 < K_{s_0} \leq 13.0$}    & \multirow{2}{*}{bluer-bRC} & \multirow{2}{*}{redder-bRC}  \\ 
(bRC)                           & & \\
\hline
{$13.0 < K_{s_0} \leq 13.7$}    & \multirow{2}{*}{bluer-fRC} & \multirow{2}{*}{redder-fRC}  \\  
(fRC)                           & & \\
\hline
$K_{s_0}\geq13.7$               & \multicolumn{2}{c}{faint RGB (fRGB)}                             \\                    
\hline
\label{Tab2}
\end{tabular}
\end{table}
%%%%%%%%%%%%%%%%%%%%%%%%%%%%%%%%%%%%%%%%%%%%%%%%%%%%%%%%%%%%%%%%%%%%%

%%%%%%%%%%%%%%%%%%%%%%%%%%%%%%%%%%%%%%%%%%%%%%%%%%%%%%%%%%%%%%%%%%%%%
\begin{table}
%\scriptsize
\setlength{\tabcolsep}{7pt}
\caption{Index Measurements for the final sample stars}
\centering         
%\begin{tabular}{clll}
\begin{tabular}{p{1.2cm}p{0.8cm}p{1.5cm}p{3.5cm}} 
\hline\hline 
Col. Number & Units & Label     & Explanations                                  \\
\hline
1           & --    & ID        & Identifier (Observation for this study)       \\
2           & deg   & RAdeg     & Right Ascension in decimal degrees (J2000)    \\
3           & deg   & DEdeg     & Declination in decimal degrees (J2000)        \\
4           & --    & field     & Field name                                    \\
5           & mag   & K0mag     & Extinction corrected 2MASS $K_s$-band magnitude  \\
6           & --    & S/N       & Signal-to-Noise ratio at 4000A                \\
7           & --    & CN        & The CN4142 index                              \\
8           & --    & errCN     & Uncertainty in CN4142 index                   \\
9           & --    & deltaCN   & Difference in CN4142 index from fiducial line \\
10          & --    & CH        & The CH4300 index                              \\
11          & --    & errCH     & Uncertainty in CH4300 index                   \\
12          & --    & deltaCH   & Difference in CH4300 index from fiducial line \\
13          & --    & CaHK      & The CaHK index                                \\
14          & --    & errCaHK   & Uncertainty in CaHK index                     \\
15          & --    & deltaCaHK & Difference in CaHK index from fiducial line   \\
16          & --    & Group     & Group type                                    \\
\hline
\label{Tab3}
\end{tabular}
\vspace{0.5mm}
\parbox{\textwidth}{\footnotesize (This table is available in its entirety in machine-readable form.)}
\end{table}
%%%%%%%%%%%%%%%%%%%%%%%%%%%%%%%%%%%%%%%%%%%%%%%%%%%%%%%%%%%%%%%%%%%%%

%%% Figure 7 %%%%%%%%%%%%%%%%%%%%%%%%%%%%%%%%%%%%%%%%%%%%%%%%%%%%%%%%
\begin{figure}
\centering
   \includegraphics[width=0.45\textwidth]{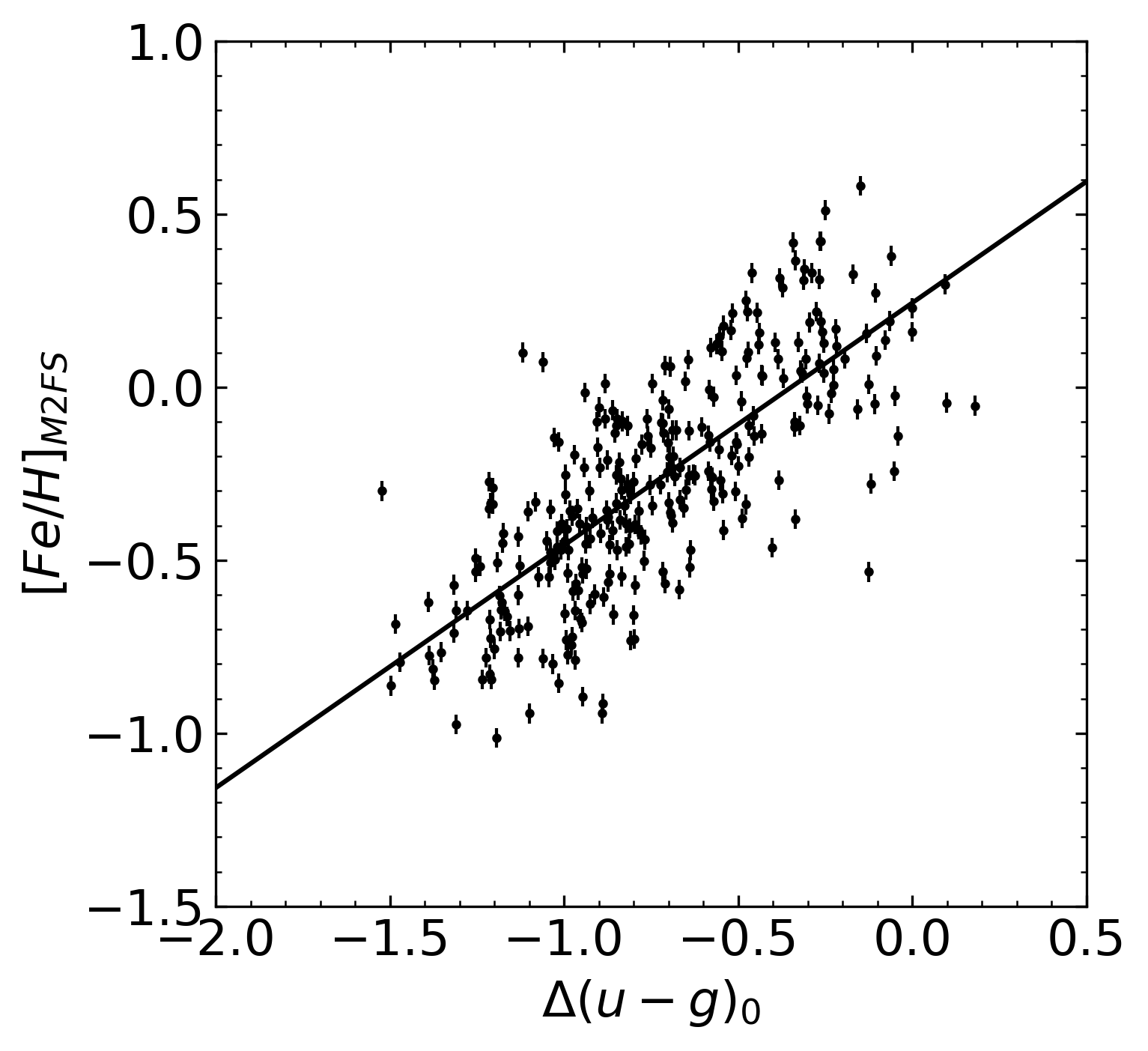} 
     \caption{
     Relation between $\Delta(u-g)_0$ color and [Fe/H] obtained from high-resolution spectroscopy using M2FS. The [Fe/H] values and their errors are taken from \citet{Lim2021a}. The black solid line represents the least-square fitting. The standard deviation of the offset between the spectroscopic [Fe/H] and the least-square fitting line is $\sim$0.21 dex.}
     \label{Fig7}
\end{figure}
%%%%%%%%%%%%%%%%%%%%%%%%%%%%%%%%%%%%%%%%%%%%%%%%%%%%%%%%%%%%%%%%%%%%%

%%% Figure 8 %%%%%%%%%%%%%%%%%%%%%%%%%%%%%%%%%%%%%%%%%%%%%%%%%%%%%%%%
\begin{figure}
\centering
   \includegraphics[width=0.3\textwidth]{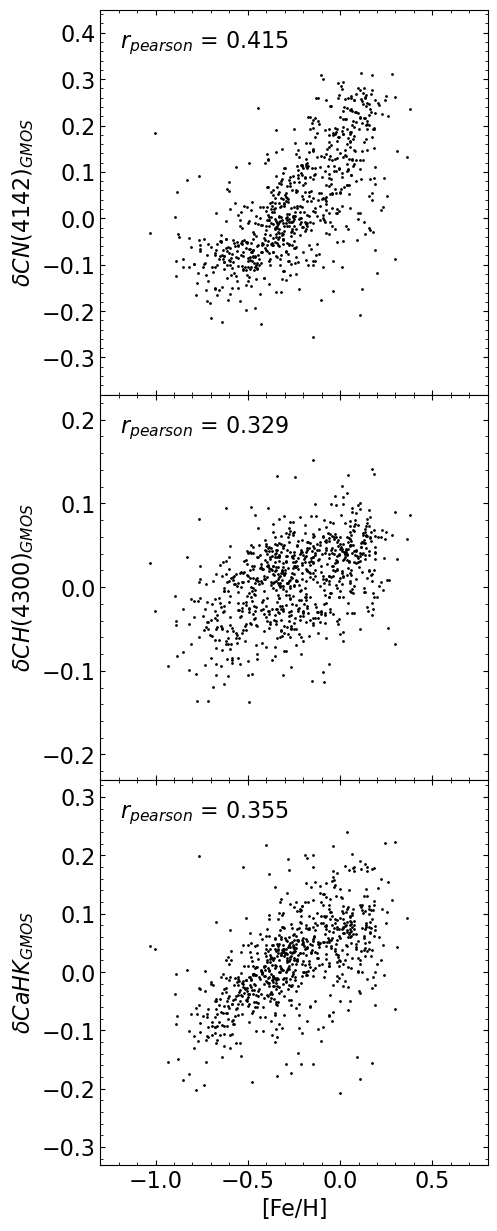} 
     \caption{Relations between the photometric [Fe/H] from the color-metallicity relation and spectral indices obtained from GMOS-S. Every indices show good correlations with photometric [Fe/H]. The Pearson correlation coefficients are indicated in the upper left corners.}
     \label{Fig8}
\end{figure}
%%%%%%%%%%%%%%%%%%%%%%%%%%%%%%%%%%%%%%%%%%%%%%%%%%%%%%%%%%%%%%%%%%%%%

\section{Relations between metallicity and spectral indices} \label{sec:result1}
\citet{Johnson2020} and \citet{Lim2021b} derived the color-metallicity relations with the $(u-g)_0$ and $(u-i)_0$ colors for the stars in the bulge fields. In this study, we used $\Delta(u-g)_0$ color for the analysis in order to minimize the bias caused by a general trend between the colors and magnitudes of RGB stars as described in Section~\ref{subsec:RCsubgrouping}. Therefore, we obtained the color-metallicity relation based on $\Delta(u-g)_0$ as:
\begin{small}
\begin{gather*}
	[Fe/H] = (0.701 \pm 0.035) \times \Delta(u-g)_0 + (0.243 \pm 0.028).
\end{gather*}
\end{small}
This relation is derived from the least-square fitting for a total number of 304 M2FS sample stars, cross-matched with BDBS data (see Figure~\ref{Fig7}). The standard deviation of the offset between the spectroscopic [Fe/H] and the least-square fitting line is $\sim$0.21 dex. The trend and the offset are generally agree with those obtained from previous studies (\citealt{Johnson2020,Lim2021b}). According to the updated relation, [Fe/H] $\sim-$0.1 dex corresponds to $\Delta(u-g)_0\sim-0.5$, which is the criteria used for dividing the bluer and redder RC groups (see Section~\ref{subsec:RCsubgrouping}). We note that the [Fe/H] of $\sim-0.1$ is comparable to the values dividing bimodal metallicity distribution function of the bulge in previous studies (e.g., \citealt{Zoccali2017}). Therefore, our subgrouping based on the $\Delta(u-g)_0$ color in the previous section correlates with two metallicity groups in the bulge, where the bluer RC represent a metal-poor population, and the redder RC indicates a metal-rich population.

As shown in Figure~\ref{Fig5}, the stars in the bRC and fRC show the different distributions in $\Delta(u-g)_0$ at a given bulge field, and the difference changes with $l$ and $b$. The difference in the median values of the $\Delta(u-g)_0$ between the bRC and fRC are approximately 0.127 for m1m8.5, 0.169 for m1m7, and 0.133 for p1m7 fields. Based on the relation between $\Delta(u-g)_0$ and [Fe/H], we could estimate the difference in [Fe/H] of 0.09, 0.12, 0.09 dex for m1m8.5, m1m7, and p1m7 fields, respectively. The estimated value of difference in [Fe/H] for m1m8.5 field is smaller than the observed value (0.149 dex; \citealt{Lim2021a}), which is probably due to the different sample size between spectroscopic ($N_{spec}\sim200$) and photometric ($N_{phot}\sim6000$) data.

As the presence of metallicity difference between bRC and fRC is further supported by various observations, comparing spectral indices between different RC groups should be conducted more carefully, taking into account the metallicity effect. Firstly, we estimated photometric [Fe/H] for all sample stars observed with GMOS-S using the color-metallicity relation. Then we compared this [Fe/H] with the spectral indices in Figure~\ref{Fig8}. The CN(4142), CH(4300), and CaHK indices are well correlated with the [Fe/H], with Pearson correlation coefficients of 0.415, 0.329, and 0.355, respectively. These relations indicate that those indices are strongly affected by metallicity. Among the specral indices, the CN index shows the most tight correlations with [Fe/H] derived from $\Delta(u-g)_0$ color, which is probably because the CN band is one of the strongest absorption lines in the BDBS $u$-filter\footnote{We note that the CN band that lies within throughput of the $u$-filter is not CN(4142), but CN(3839). Nevertheless, the strengths of them obviously trace the same species, the CN molecule, and correlated each other.}. These trends further imply that the variations in $\delta$CN, $\delta$CH, and $\delta$CaHK indices among stars could only be attributed to the variations in N, C, and Ca abundances, respectively, when the [Fe/H] of the stars are the same, such as the GC environment. Thus, we need to revisit the difference in CN band strength between the bRC and fRC reported by \citetalias{Lee2018}.

\citetalias{Lee2018} interpreted the stronger CN band strength in the bRC compared to the fRC as a result of N enhancement in the bRC, on the basis of similar distribution of [Fe/H] for the two RC groups (\citealt{DePropris2011,Uttenthaler2012}). Since, however, a significant [Fe/H] difference between the bRC and fRC are reported (0.149 dex for m1m8.5 field; \citealt{Lim2021a}), the effect of metallicity difference can no longer be neglected when explaining the difference in CN band strength between the two RCs in the bulge. As shown in the top panel of Figure~\ref{Fig8}, a 0.15 dex difference in of [Fe/H] corresponds to roughly a 0.05 difference in the $\delta$CN index, which is comparable to the value reported by \citetalias{Lee2018} ($\sim$0.052; see Figure 4(d) of \citetalias{Lee2018}). Therefore, it is hard to conclude whether the difference in CN band strength is primarily due to N enhancement or the difference in metallicity solely by comparing the mean values of the bRC and fRC. In order to constrain the metallicity of the RC stars, we subgrouped the bRC and fRC stars into bluer (metal-poor) and redder (metal-rich) groups based on the $\Delta(u-g)_0$ color (see Table~\ref{Tab2}). The comparison of the spectral indices between the bright and faint RC groups then made within the similar metallicity ranges, which is presented in the forthcoming section.

%%% Figure 9 %%%%%%%%%%%%%%%%%%%%%%%%%%%%%%%%%%%%%%%%%%%%%%%%%%%%%%%%
\begin{figure*}
\centering
   \includegraphics[width=0.7\textwidth]{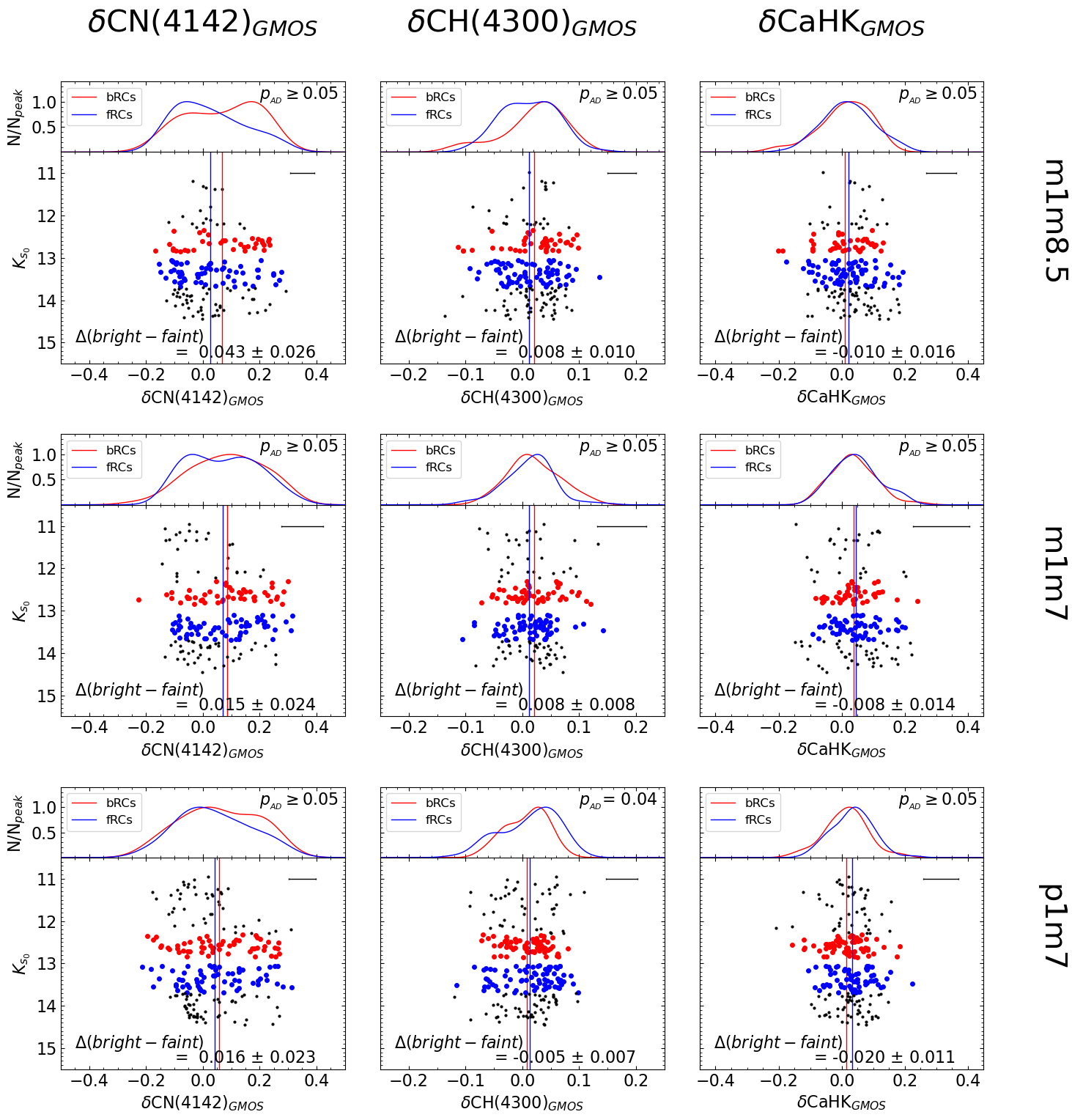} 
     \caption{Top row: Measured $\delta$CN(4142) (left), $\delta$CH(4300) (middle), and $\delta$CaHK (right) indices for stars in m1m8.5 field observed with GMOS-S. The color code is the same as in Figure~\ref{Fig2}. The blue and red vertical lines denote the mean value for each RC group. The horizontal bar in the upper right corner indicates the typical measurement error. The gaussian KDE for stars in bRC and fRC regimes are also shown in top panels with $p$-values derived from AD test for the two distributions. Middle and bottom rows: Same as the top row, but for the m1m7 and p1m7 fields, respectively. %The data used to create this figure are available.
     }    
     \label{Fig9}
\end{figure*}
%%%%%%%%%%%%%%%%%%%%%%%%%%%%%%%%%%%%%%%%%%%%%%%%%%%%%%%%%%%%%%%%%%%%%

\section{Difference in chemical properties between red clump subgroups} \label{sec:result2&3}
\subsection{`Bright RC' versus `Faint RC'} \label{subsec:result2}
We first compared the spectral indices between the bRC and fRC for the three fields in Figure~\ref{Fig9}. For $\delta$CN(4142) index, the mean difference between the stars in the two RC regimes at m1m8.5 field is measured to be $0.043\pm0.026$ mag, where the uncertainty is estimated based on the error of the mean of each group. This value is comparable to that obtained at the same field in \citetalias{Lee2018} ($0.053\pm0.011$ mag). The differences in $\delta$CN(4142) at m1m7 and p1m7 fields are, respectively, $0.015\pm0.024$ and $0.016\pm0.023$ mag, which are smaller than that at m1m8.5 field. The $p$-values obtained from the Anderson-Darling (AD) test for the three fields are not sufficient to reject the null hypothesis that the bRC and fRC groups are drawn from the same population ($p_{_{AD}}\geq0.05$ for all three fields). Although these differences are statistically less significant, the shape of the distribution for each group, presented in the upper panels of Figure~\ref{Fig9}, is similar to that of the $\Delta(u-g)_0$ color distribution in Figure~\ref{Fig5}. This similarity is possibly because the CN band is one of the strongest features in the BDBS $u$-band region as discussed in Section~\ref{sec:result1}. 

Although the $p$-value of the $\delta$CH(4300) at p1m7 field is a little bit less than 0.05, the mean of $\delta$CH(4300) and $\delta$CaHK show negligible differences between the bRC and fRC for all fields, which is consistent with results of \citetalias{Lee2018} for the m1m8.5 field. The differences are less than 0.02 mag, which are comparable to those uncertainties. 

%%% Figure 10 %%%%%%%%%%%%%%%%%%%%%%%%%%%%%%%%%%%%%%%%%%%%%%%%%%%%%%%%
\begin{figure}
\centering
   \includegraphics[width=0.45\textwidth]{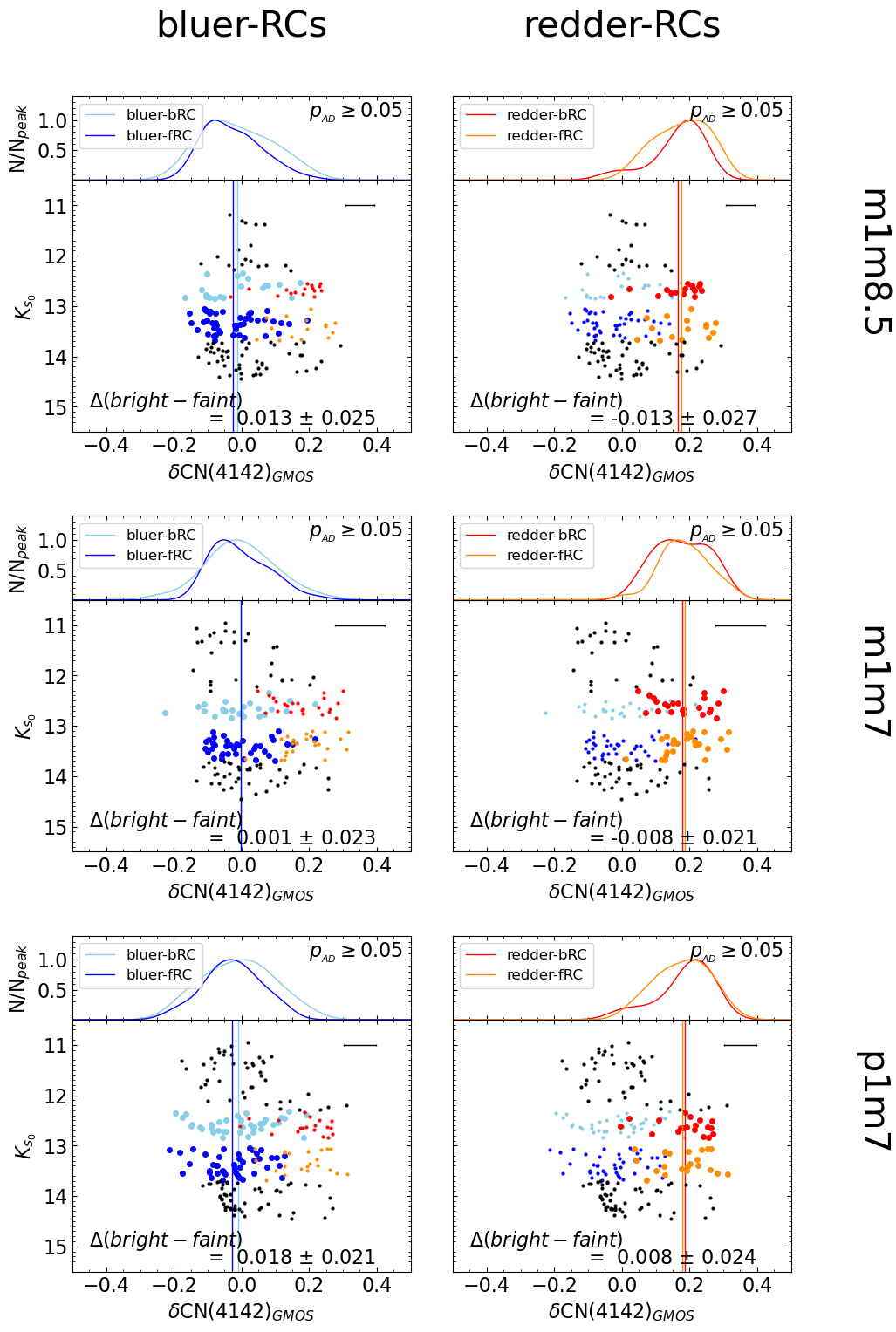} 
     \caption{Left column: Comparisons of distributions of $\delta$CN(4142) obtained from GMOS-S between the bluer-bRC (lightblue) and bluer-fRC (darkblue) subgroups. The RGB (black), redder-bRC (red), and redder-fRC (orange) are also presented with smaller circles. The solid straight lines denote the mean value for each group, and the gaussian KDE for the two RC groups are also shown in top panels with $p$-values. Right column: Same as the left column, but between the redder-bRC (red) and redder-fRC (orange). Note that the differences in $\delta$CN(4142) between the bright and faint groups are not statistically significant for both the redder and the bluer stars.}
     \label{Fig10}
\end{figure}
%%%%%%%%%%%%%%%%%%%%%%%%%%%%%%%%%%%%%%%%%%%%%%%%%%%%%%%%%%%%%%%%%%%%%

%%% Figure 11 %%%%%%%%%%%%%%%%%%%%%%%%%%%%%%%%%%%%%%%%%%%%%%%%%%%%%%%%
\begin{figure}
\centering
   \includegraphics[width=0.45\textwidth]{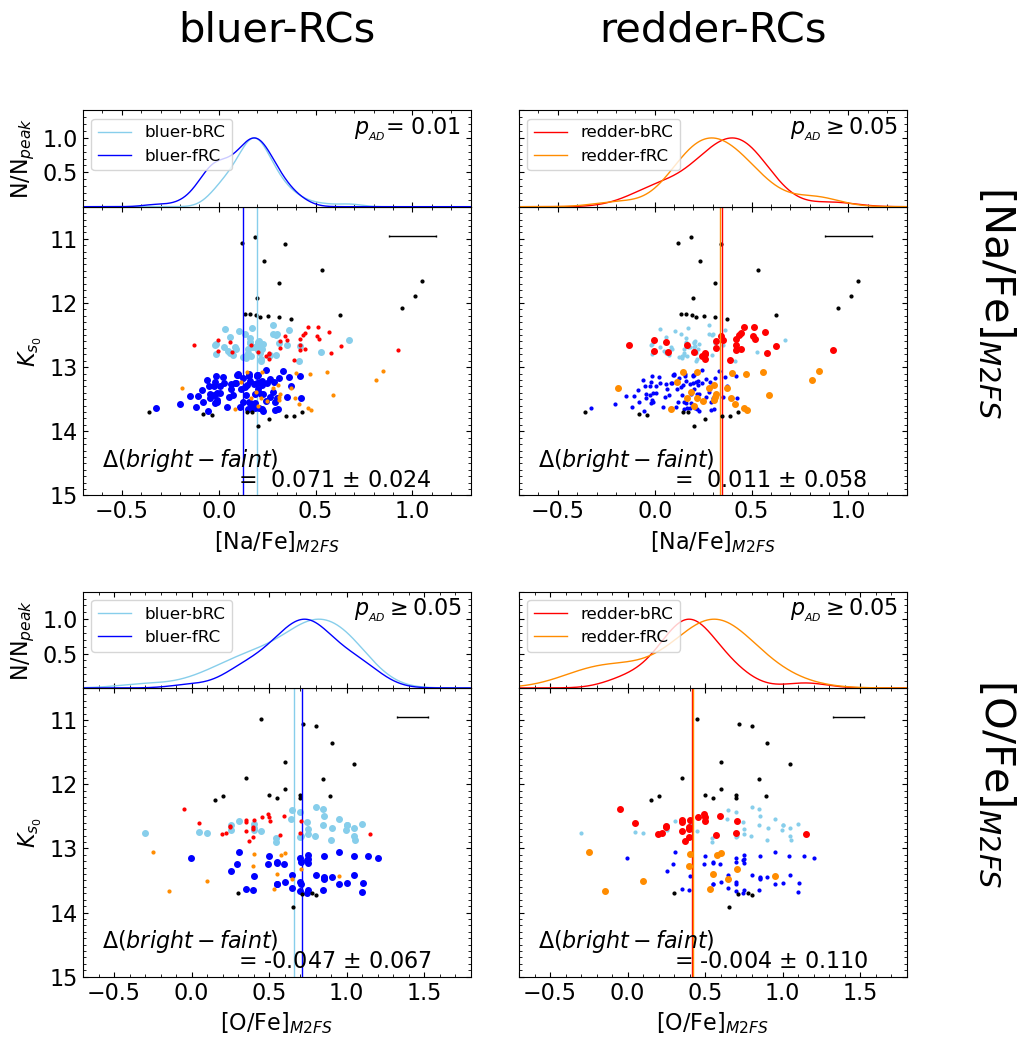} 
     \caption{Comparisons of distributions of [Na/Fe] (top row) and [O/Fe] (bottom row) obtained from high-resolution spectroscopy (\citealt{Lim2021a}) between RC subgroups for m1m8.5 field. The comparisons between the bluer-bRC (lightblue) and bluer-fRC (darkblue) are shown in the left column, and those between the redder-bRC (red) and redder-fRC (orange) are presented in the right column. The difference in [Na/Fe] between the bluer-bRC and bluer-fRC is 0.071 dex with a confidence level of 2.96$\sigma$, while no statistically significant differences in Na is observed between the redder-bRC (red) and redder-fRC (orange) groups.}
     \label{Fig11}
\end{figure}
%%%%%%%%%%%%%%%%%%%%%%%%%%%%%%%%%%%%%%%%%%%%%%%%%%%%%%%%%%%%%%%%%%%%%

\subsection{'Bright RC' versus 'Faint RC' within the same $u$-band color regimes} \label{subsec:result3}
As described in Section~\ref{subsec:RCsubgrouping}, each of the bRC and fRC groups are further divided into bluer and redder groups. Thus, the comparison of spectral indices in similar $\Delta(u-g)_0$ color ranges is required. Figure~\ref{Fig10} shows the comparisons of the $\delta$CN(4142) index between the stars in the bluer-bRC and bluer-fRC, and those between the redder-bRC and redder-fRC stars at the three fields. The redder stars have higher mean value of $\delta$CN(4142) compared to that of bluer stars, which is because the redder stars are more metal-rich, and the $\delta$CN(4142) is strongly correlated with the metallicity (see Figure~\ref{Fig8}). For all three fields, no statistically significant difference in $\delta$CN(4142) is observed between the bluer-bRC and bluer-fRC, as well as between the redder-bRC and redder-fRC. Similar results are obtained from $\delta$CH(4300) and $\delta$CaHK indices. Since the X-shaped bulge model predicts the identical chemical composition of bright and faint RC stars, the results appear to support the X-shaped bulge scenario as the origin of the double RC for both metal-poor and metal-rich populations. 

However, it is still possible that RC stars in the brighter groups would be enhanced in N, which are not detectable by measuring of the CN band strength in this study. As we described in Section~\ref{indexmeasurement}, we employed CN(4142) index instead of CN(3839), because the later index suffer from saturation at the bRC and bRGB regimes. However, the CN(4142) index is known to have less sensitivity to N abundance and higher error compared to the CN(3839) index (\citealt{Harbeck2003,Pancino2010}). In addition, insufficient number of the sample stars in this study and heavy contamination of the RGB stars on the RC regimes would make it more difficult to detect the possible difference in N abundance (see Section~\ref{sec:Discussion&Conclusion} for further discussion).

Interestingly, unlike the results of the CN index, we found a sign of Na enhancement for the stars in the bluer-bRC compared to those in the bluer-fRC at m1m8.5 field (see top-left panel of Figure~\ref{Fig11}). The element abundances for the stars at the m1m8.5 field are obtained from high-resolution spectroscopy of \citet{Lim2021a}. The bluer-bRC stars show an enhancement in [Na/Fe] by 0.071$\pm$0.024 dex\footnote{\citet{Lim2021a} reported 0.053$\pm$0.021 dex difference in [Na/Fe] between stars in the `bluer+redder' bRC and `bluer+redder' fRC regimes. In our sample criteria, the difference in [Na/Fe] between the `bluer+redder' bRC and `bluer+redder' fRC groups is measured to be 0.076$\pm$0.027 dex.} compared to the bluer-fRC stars. When considering the uncertainty of the difference which is obtained based on the error of the mean of each group, the level of confidence of the difference is estimated to be 2.96$\sigma$. A two-sample AD test also indicates that the two groups are likely drawn from the different distribution with a $p$-value of 0.01. We suspect that the difference in sample size is one of the reasons that cause statistically significant difference in Na but negligible difference in CN between the two groups. We note that 135 stars were used in the comparison of Na abundances between the bluer-bRC and bluer-fRC (top-left panel of Figure~\ref{Fig11}), whereas only 64 stars were usable in the comparison of CN index (top-left panel of Figure~\ref{Fig10}).

The observed Na difference between the bluer-bRC and bluer-fRC is, however, not the difference between the genuine RC stars in the two groups. Because the RC zones in CMD are mixed with background RGB stars, the difference in [Na/Fe] caused by the genuine RC stars would be diluted in the observations of this study. In Figure~\ref{Fig2}, the relative fraction of RGB stars in the bRC and fRC zone is estimated to be 66\% and 78\%, respectively, from the fitted luminosity functions with and without stars in RC zone\footnote{The regions of red clump stars also include the RGB bump stars, which would make the relative fraction of RGB stars even higher than presented values. This point is discussed in Section 6.2}. When the contributions of the RGB stars on the RC regimes are taken into account, the difference between the genuine bluer RC stars would correspond to $\Delta$[Na/Fe]$\sim$0.23 dex, which is comparable to that between G2+ and G1 stars observed in GCs (\citealt{Carretta2009a}). Little or no difference in [O/Fe] is observed in Figure~\ref{Fig11} between the bluer-bRC and bluer-fRC, unlike to GCs with multiple populations showing Na-O anti-correlation (\citealt{Carretta2009a,Bastian2018}). However, the absence of O-depletion among RC stars can be understood in the multiple population scenario for the double RC, because a significant depletion is only observed in the most extreme subpopulation in GCs, while the intermediate subpopulations shows minor depletion (see, e.g., \citealt{Carretta2009a,Carretta2015}). In this regard, it is significantly more challenging to observe the potential O-depletion of bluer-bRC. In the case of redder stars, on the other hand, no statistically meaningful difference in Na, as well as O, is observed between the redder-bRC and redder-fRC stars, indicating no evidence of multiple populations, unlike to the bluer stars.

Although further observations for the Na abundance measurements are required, the different chemical patterns of bluer and redder RCs indicates the different origin of the double RC: bluer stars are likely originated from multiple populations, whereas redder stars are more likely from X-shaped structure.

\section{Discussion and conclusion} \label{sec:Discussion&Conclusion}
In order to investigate the origin of the double RC phenomenon observed in the CMD of the MW bulge, we performed low-resolution spectroscopy for the RC and RGB stars in three fields of the bulge. When we subdivide each of the bRC and fRC into bluer and redder subgroups based on the $\Delta(u-g)_0$ color, no statistically meaningful differences in CN have been detected between the bright and faint groups for both bluer and redder regimes in all three fields. Despite of the negligible difference in CN, we found an enhancement of [Na/Fe] in the bluer-bRC compared to the bluer-fRC in the m1m8.5 field from the cross-matching with the high-resolution spectroscopic data. Whereas, the redder-bRC and redder-fRC groups do not show any meaningful difference in chemical properties. Our results indicate that the origins of double RC phenomenon could be different in the bluer (metal-poor) and redder (metal-rich) populations, i.e., multiple population origin of the double RC in the metal-poor stars, and the distance effect caused by X-shaped structure in the metal-rich component. It implies that our MW bulge is composed of a metal-poor classical bulge component and a metal-rich pseudo bulge component with an X-shaped structure, which is consistent with the previous studies (\citealt{Rojas-Arriagada2014,Zoccali2018,Kunder2020,Lim2021b}).

\subsection{Impact of the metallicity on the spectral indices} \label{chapter3:discussion:limitation}
In the study of GCs, the CN and CH bands can be used as effective tracers of N and C abundances because cluster member stars have almost similar metallicity (see, e.g., \citealt{smith1996,simpson2017}). Unlike the studies for the GC, it is necessary to consider the metallicity effect on the strength of the CN and CH bands in studies of Galactic halo and bulge. For example, in order to estimate the contribution of GCs to the formation of the halo, \citet{Martell2011} divided the halo stars into different metallicity bins, and labeled the stars showing strong CN and weak CH band strengths as GC-originated second generation stars at fixed metallicity bin (see also \citealt{Martell2010,Koch2019}). In our previous study for the bulge stars (\citetalias{Lee2018}), the CN band has also been used to verify N abundance difference between the bRC and fRC based on the previous knowledge that the bRC and fRC have identical chemistry in average (\citealt{DePropris2011,Uttenthaler2012}). Since, however, a statistically significant difference in [Fe/H] between the two groups has been discovered by \citet{Lim2021a}, the difference in the mean of the CN band strength reported by \citetalias{Lee2018} no longer only represent the difference in N abundance, but the metallicity effect should be considered. In order to constrain the metallicity of the RC stars, in this study, we subdivided the RC stars into bluer and redder groups based on the $\Delta(u-g)_0$ color, and then compared CN band strength between the bright and faint RC groups within the similar metallicity regime. We note, however, that the $\Delta(u-g)_0$ color is still incomplete to constrain metallicity effect independent of the N enhancement, because the CN molecular band is one of the strongest absorptions in the BDBS $u$-filter. Therefore, other metallicity indicators independent from the CN band are required to trace N enhancement in the metal-complex system, like the bulge. We expect that narrow-band photometric survey, similar to the Pristine survey (\citealt{Starkenburg2017}), for the bulge region could provide a better constraint on the metallicity of the bulge stars.

\subsection{Impact of RGB stars on red clump populations} \label{chapter3:discussion:rgb}
The RC regions in CMD of the bulge are not only composed of genuine RC stars, but also mixed with the RGB stars which are indistinguishable from the RC stars by spectroscopy (\citealt{Boberg2016,Masseron2017}). Therefore, as described in Section~\ref{subsec:result3}, the difference in chemical properties between the genuine RC stars in different magnitude ranges would be diluted in the observations of this study. When estimating the contribution of RGB stars on the RC regions, the effect of RGB bump (RGBB) also needs to be considered because the RGBB corresponding to the bright RC population would be placed near faint RC magnitude range (see Figure 3 of \citealt{Nataf2013b}). In addition, \citet{Nataf2013a} showed from the photometric observations for the MW GCs that the relative number of RGBB stars to RC stars increases at higher-metallicity. For example, the relative number of RGBB stars to RC stars ($N_{RGBB}/N_{RC}$) in GCs at [M/H]$\sim$0 is approximately 30\% in Figure 11 of \citet{Nataf2013a}. The fraction of RGB and RC stars in the fRC zone is estimated to be 0.78:0.22 (see Figure~\ref{Fig2}). Therefore, the expected fraction of genuine RC stars to the total number of stars in the fRC regime would only be 15.4\% (22\% $\times$ 70\%) when the contribution of RGBB stars is roughly taken into account. This small fraction would make it more difficult to detect the possible abundance difference between the genuine RC stars.

Due to the serious impact of RGB stars on the RC populations, spectroscopy based on the random target sampling in the RC regions would be inefficient to detect possible chemical abundance difference between the bright and faint RC groups expected in the multiple population scenario. We note that it is possible to distinguish RC stars from the RGB stars by asteroseismology for the nearby stars (\citealt{Bedding2011,Mosser2014}), but stars in the bulge are too distant to obtain meaningful asteroseismic data with current space-based facilities. If it becomes possible for the bulge stars in the near future with Roman Space Telescope (\citealt{Huber2023}), the RC spectroscopic survey with RC-only-sampling is expected to be available.

\subsection{Conclusion}
It is now obvious that each of the bright and faint RC group is not a single group, but consists of two distinct subgroups with different metallicity (\citealt{Johnson2020,Lim2021b,Johnson2022}). We suggest, based on our results, different formation origins of the the metal-poor and metal-rich components in the Galactic bulge as below:

\textit{Metal-poor ([Fe/H]$<$-0.1) stars}: We found that the stars in the bluer-bRC group show a sign of Na enhancement compared to those in the bluer-fRC group (Section~\ref{subsec:result3}). The difference in [Na/Fe] between the two RC populations is estimated to be $\sim$0.23 dex when the contribution of RGB stars on the RC regimes is taken into account, which is comparable to that between G2+ and G1 stars observed in MW GCs. It is now well established that GCs are the only environment in which the Na enhanced populations can be produced, and the Na enhanced G2+ stars are also enhanced in N and He in GCs with multiple populations (\citealt{Gratton2012,Renzini2015,Bastian2018}). Therefore, the Na enhancement in the bluer-bRC stars implies that the double RC observed in the metal-poor stars is likely due to the multiple populations originated from GCs. A strong bimodal distribution in Na abundance of stars at the outer bulge field reported by \citet{Lee2019} is aligned with our result, and further strengthen the multiple population origin of the double RC. One possible candidates which have provided both the N/Na/He-enhanced G2+ stars and G1 stars with normal abundances to the bulge field are the primordial GCs that already disrupted. The presence of N-enhanced stars in the inner bulge (\citealt{Schiavon2017}), and the existence of N-enriched galaxies at high redshift (\citealt{Charbonnel2023,Senchyna2024}) also support this hypothesis that the primordial GCs have played a role in the formation of the Galactic bulge. Indeed, the MW GCs Terzan 5 and Liller 1, which are located in the bulge and host multiple stellar populations, have been revealed as survived remnants of building blocks of the bulge formation (\citealt{Ferraro2016,Ferraro2021}). The detailed link between their stellar population properties and the formation of the Galactic bulge is, however, remained to be answered. Recent studies with JWST observations suggest a possible connection between the primordial GCs and the stellar clumps in high-z galaxies (\citealt{Claeyssens2023,Vanzella2023}). Although hydrodynamic simulations considering star forming clumps in the early gas-rich disk well reproduce the chemical imprints observed in the MW bulge (\citealt{Debattista2023,Garver2023}), more detailed chemical evolution model including N and Na abundances is required to be included in the simulations for the clumpy galaxies.

\textit{Metal-rich ([Fe/H]$\geq$-0.1) stars}: Our results show no obvious difference in chemical compositions between the redder-bRC and redder-fRC groups, which is consistent with the expectation of the X-shaped bulge scenario. In addition to our results, the variations of the number ratio between the redder-bRC and redder-fRC groups on the $(u-g,~K_s)_0$ CMDs in various fields of the bulge (\citealt{Lim2021b}), and cylindrical rotations observed in the metal-rich stars (\citealt{Kunder2012,Ness2013,Zoccali2014}) are also consistent with the characteristic of the X-shaped pseudo bulge. However, the current modelling of the X-shaped structure of the MW bulge have been constructed without considering the discrete distribution of RC stars in the metallicity (see, e.g., \citealt{Wegg2013}). Therefore, in addition to the split of RC stars in magnitude, the number distribution of RC stars in the metallicity-sensitive color should be considered for the detailed modelling of the 3D structure of the Galactic bulge.

Our results are in support of the composite bulge model (\citealt{Rojas-Arriagada2014,Zoccali2018,Kunder2020,Lim2021b}). The metal-rich component with the X-shaped structure is likely the product of the bar-driven secular evolution of the inner disk (\citealt{DiMatteo2016,Debattista2017}), whereas the metal-poor spheroidal component would have formed by the assembly of GC-like subsystems, in the form of a proto-GCs or star forming clumps in the ``clump-origin bulge" scenarios (\citealt{Noguchi1999,Elmegreen2008,Inoue2012,Pfeffer2018,Lee2019}). For better understanding of the formation and evolution of the Galactic bulge, diverse approaches are essential, including dedicated spectroscopic observations for the stars in the bulge fields aimed at measuring key elements to trace second generation stars, hydrodynamic simulations for the clumpy galaxies including detailed chemical evolution model, and dynamical modelling for the metal-poor and metal-rich components of the bulge.

%% Also note that the akcnowlodgment environment does not support long amounts of text. If you have a lot of people and institutions to acknowledge, do not use this command. Instead, create a new \section{Acknowledgments}.
\begin{acknowledgments}
We are greatful to the referee and the statistics editor for a number of helpful suggestions. We also thank Christian Johnson for providing BDBS data. 
Support for this work was provided by the National Research Foundation of Korea to the Center for Galaxy Evolution Research (2022R1A6A1A03053472 and 2022R1A2C3002992). 
This work was supported by the LAMP Program of the National Research Foundation of Korea (NRF) grant funded by the Ministry of Education (No. RS-2023-00301976).
This work was supported by K-GMT Science Program (PID: GS-2018B-Q-211, GS-2019A-Q-118, GS-2019A-Q-223, GS-2020B-Q-115, and GS-2021A-Q-215) of Korea Astronomy and Space Science Institute (KASI).
S.H. thanks Jubee Sohn for comments and encouragements. 
\end{acknowledgments}




\bibliography{Hong_et_al_2024_ApJ_accepted.bib}{}
\bibliographystyle{aasjournal}

\end{document}